\documentclass[a4paper]{scrartcl}
\pdfoutput=1
\usepackage[utf8]{inputenc}

\usepackage[T1]{fontenc} 
\usepackage{lmodern} 
\usepackage[sc,osf]{mathpazo} 
\AtBeginDocument{%
  \DeclareFontShape{T1}{pplj}{m}{scit}{<-> ec-qplri-sc}{}
}
\usepackage{myT1classico} 
\DeclareMathAlphabet{\mathsf}{T1}
  {\sfdefault}{m}{n} 
\SetMathAlphabet{\mathsf}{bold}{T1}{\sfdefault}{b}{n} 

\usepackage{amsmath,amssymb,amsthm}
\usepackage{mathtools}
\usepackage{tensor}

\usepackage{accsupp}
\newcommand*{\acronym}[1]{\texorpdfstring{%
    \protect\BeginAccSupp{%
      method=pdfstringdef,%
      ActualText=#1
    }%
    \textsc{\textls[40]{\MakeLowercase{#1}}}%
    \protect\EndAccSupp{}}%
  {#1}%
}

\usepackage[british]{babel}

\usepackage{microtype} 

\usepackage{authblk}

\usepackage[colorlinks]{hyperref}
\usepackage[nameinlink]{cleveref}

\DeclareFontFamily{U}{matha}{\hyphenchar\font45}
\DeclareFontShape{U}{matha}{m}{n}{
  <5> <6> <7> <8> <9> <10> gen * matha
  <10.95> matha10 <12> <14.4> <17.28> <20.74> <24.88> matha12
}{}
\DeclareSymbolFont{matha}{U}{matha}{m}{n}
\DeclareMathSymbol{\oleft}{2}{matha}{"68}
\DeclareMathSymbol{\oright}{2}{matha}{"69}

\usepackage[style=phys,biblabel=brackets,eprint=true,sorting=noney,
  sortcites=false]{biblatex}
\newbibmacro{string+doiurl}[1]{%
  \iffieldundef{doi}{%
    \iffieldundef{url}{#1}{\href{\thefield{url}}{#1}}%
  }{\href{https://doi.org/\thefield{doi}}{#1}}%
}
\DeclareFieldFormat{title}{\usebibmacro{string+doiurl}{\mkbibemph{#1}}}
\DeclareSortingTemplate{noney}{
  \sort{\citeorder}
}

\DefineBibliographyExtras{british}{}

\newcommand*{\e}{\mathrm{e}} 
\newcommand*{\I}{\mathrm{i}} 
\newcommand*{\D}{\mathrm{d}} 
\newcommand*{\R}{\mathbb R}
\newcommand*{\Gal}{\mathsf{Gal}}
\newcommand*{\bomega}{{\boldsymbol{\omega}}}
\newcommand*{\ba}{{\boldsymbol{a}}}
\newcommand*{\btheta}{{\boldsymbol{\theta}}}
\newcommand*{\Ui}{\mathsf{U}(1)}
\newcommand*{\ui}{\mathfrak{u}(1)}

\theoremstyle{plain}
\newtheorem{theorem}{Theorem}
\newtheorem{corollary}[theorem]{Corollary}

\makeatletter
\mathchardef\ordinarycolon\mathcode`\:%
\mathcode`\:=\string"8000%
\begingroup \catcode`\:=\active%
  \gdef:{\@ifnextchar{=}
    {\coloneq\@gobble}
    {\ordinarycolon}}%
\endgroup
\makeatother

\AddToHook{env/corollary/begin}{\crefalias{theorem}{corollary}}

\numberwithin{equation}{section}

\title{On gauge transformations in twistless-torsional Newton--Cartan
  geometry}

\author[1,2,3,a]{Arian L. von Blanckenburg}
\author[3,b]{Philip K. Schwartz}
\affil[1]{Max Planck Institute for Gravitational Physics (Albert
  Einstein Institute), \par
  Callinstraße 38, 30167 Hannover, Germany}
\affil[2]{Leibniz University Hannover,
  Institute of Gravitational Physics, \par
  Callinstraße 38, 30167 Hannover, Germany}
\affil[3]{Leibniz University Hannover,
  Institute of Theoretical Physics, \par
  Appelstraße 2, 30167 Hannover, Germany}
\affil[a]{\normalfont\texttt{\href{mailto:arian.von.blanckenburg@aei.mpg.de}
    {arian.von.blanckenburg@aei.mpg.de}}}
\affil[b]{\normalfont\texttt{\href{mailto:philip.schwartz@itp.uni-hannover.de}
    {philip.schwartz@itp.uni-hannover.de}}}

\date{}

\begin{document}
\maketitle

\begin{abstract}
  \noindent
  Twistless-torsional Newton--Cartan (\acronym{TTNC}) geometry exists
  in two variants, type~\acronym{I} and type~\acronym{II}, which
  differ by their gauge transformations.  In \acronym{TTNC} geometry
  there exists a specific locally Galilei-invariant function, called
  by different names in existing literature, that we dub the `locally
  Galilei-invariant potential'.  We show that in both types of
  \acronym{TTNC} geometry, there always exists a local gauge
  transformation that transforms the locally Galilei-invariant
  potential to zero.  For type~\acronym{I} \acronym{TTNC} geometry, we
  achieve this due to the corresponding equation for the gauge
  parameter taking the form of a Hamilton--Jacobi equation.  In the
  case of type~\acronym{II} \acronym{TTNC} geometry, we perform
  subleading spatial diffeomorphisms.  In both cases, our arguments
  rigorously establish the existence of the respective gauge
  transformation also in case of only finite-degree differentiability
  of the geometric fields.  This improves upon typical arguments for
  `gauge fixing' in the literature, which need analyticity.

  We consider two applications of our result.  First, it generalises a
  classical result in standard Newton--Cartan geometry.  Second, it
  allows to (locally) parametrise \acronym{TTNC} geometry in two new
  ways: either in terms of just the space metric and a unit timelike
  vector field, or in terms of the distribution of spacelike vectors
  and a positive-definite cometric.
\end{abstract}

\section{Introduction}
\label{sec:intro}

Newton--Cartan gravity \cite{Cartan:1923_24,Friedrichs:1928,
  Trautman:1963,Trautman:1965,Dombrowski.Horneffer:1964,Kuenzle:1972,
  Kuenzle:1976,Ehlers:1981a,Ehlers:1981b}, \cite[chapter~4]
{Malament:2012}, \cite{Hartong.EtAl:2023,Schwartz:NC_gravity} is a
differential-geometric reformulation of Newtonian gravity, bringing
its formulation closer to General Relativity~(\acronym{GR}).  This
allows to emphasise similarities and differences between Newtonian
gravity and \acronym{GR}.  Similar to \acronym{GR}, Newton--Cartan
gravity describes gravity and inertia by a curved connection on the
manifold of spacetime events.  However, differently to the
\acronym{GR} case, in Newton--Cartan gravity spacetime is endowed not
with a Lorentzian metric, but with a specific notion of
Galilei-relativistic metric structure, ensuring local Galilei
invariance of physical laws.  In both Newton--Cartan gravity and
\acronym{GR}, one considers connections compatible with the metric
structure.

In Newton--Cartan gravity, there are two different sensible
restrictions on the (local) notion of time provided by the metric
structures, \emph{absolute time} and \emph{absolute simultaneity}, the
first of which implies the latter.  Interestingly, the notion of time
partially restricts the torsion of the connection.  In cases where
absolute simultaneity holds, but not absolute time, compatible
connections necessarily have non-vanishing torsion.  The resulting
geometry is called twistless-torsional Newton--Cartan (\acronym{TTNC})
geometry \cite{Christensen.EtAl:2014a,Christensen.EtAl:2014b,
  Bergshoeff.EtAl:2015,Bekaert.Morand:2016,Van_den_Bleeken:2017}.
  
Already in standard Newton--Cartan geometry one can introduce, in
addition to the metric structure, an additional `auxiliary' field
related to the Bargmann algebra, the central extension of the Galilei
algebra \cite{Duval.Kuenzle:1977,Duval.Kuenzle:1984,
  Andringa.EtAl:2011,Geracie.EtAl:2015,Bekaert.Morand:2016,
  Schwartz:2023}.  In \acronym{TTNC} geometry, this field features
more heavily.  In the literature, it goes under several different
names; we will call it a `Bargmann form' (for more on this naming, see
\cref{fn:name_a} in \cref{sec:TTNC}).  There is a gauge freedom for
the Bargmann form: certain transformations of it are to be considered
gauge transformations, relating different equivalent descriptions of
the same geometric situation.  Depending on which transformations are
considered gauge, one distinguishes between so-called type~\acronym{I}
and type~\acronym{II} \acronym{TTNC} geometry \cite{Hansen.EtAl:2019,
  Hansen.EtAl:2020,Hartong.EtAl:2023}.

The Bargmann form can be used to define a locally Galilei-invariant
vector field---that is, a vector field invariantly determined by the
metric structure and the Bargmann form.  In the study of this vector
field, a certain scalar function arises naturally, going by different
names in the literature: for example, it is called `Newton potential'
in ref.~\cite{Bergshoeff.EtAl:2015} or `gravitational gauge scalar' in
ref.~\cite{Bekaert.Morand:2016}.  In this work we call it `locally
Galilei-invariant potential'.  We will show that, in both the
type~\acronym{I} and type~\acronym{II} cases, locally there exists a
gauge transformation such that after performing this transformation,
the locally Galilei-invariant potential vanishes.

This (partial) gauge fixing is interesting for two reasons.  On the
one hand, it offers a natural generalisation to \acronym{TTNC}
geometry of a classical result in standard Newton--Cartan geometry:
the local existence of twist-free geodesic unit timelike vector fields
\cite[thm.~3.6]{Dombrowski.Horneffer:1964}, \cite[prop.~4.3.7]
{Malament:2012}, \cite[prop.~3.26]{Bekaert.Morand:2016}.  On the other
hand, we will show that it allows for a (local) parametrisation of
\acronym{TTNC} geometries up to gauge by either just the space metric
and a unit timelike vector field, or the distribution of spacelike
vectors and a positive-definite cometric.

The proof of our result will be fully rigorous and apply in the case
where the involved geometric fields are only assumed finitely-often
differentiable.  We thereby improve on the existing literature by
dropping the analyticity assumption that is made for such gauge fixing
results, either explicitly (as in ref.~\cite{Bekaert.Morand:2016}) or
implicitly in typical arguments by counting parameters and equations
(as in, e.g., ref.~\cite{Van_den_Bleeken:2017}).

Throughout this paper we use the notation and conventions from
ref.~\cite{Schwartz:2023}.  In particular, the local representative of
the Bargmann form is called $a$ and not $m$ as in some other
literature on \acronym{TTNC} geometry.\footnote{The literature which
  is more quantum-field-theory-adjacent usually uses $m$, the more
  relativity-adjacent literature uses $a$ or $A$.}

The structure of this paper is as follows.  In \cref{sec:TTNC} we give
a brief review of \acronym{TTNC} geometry.  We show how local gauge
transformations can be used to `gauge away' the locally
Galilei-invariant potential in both type~\acronym{I} and
type~\acronym{II} \acronym{TTNC} geometry in
\cref{sec:gauge_phi_hat_0}.  In \cref{sec:application} we apply this
to parametrisation of \acronym{TTNC} geometries as mentioned above,
and explain how it generalises the classical result.  We conclude and
discuss further research directions in \cref{sec:conclusion}.

\section{Twistless-torsional Newton--Cartan geometry}
\label{sec:TTNC}

In this section, we briefly review \acronym{TTNC} geometry in a style
similar to that of references~\cite{Kuenzle:1976,Ehlers:1981a,
  Ehlers:1981b}, \cite[chapter~4]{Malament:2012}, \cite{Schwartz:2023,
  Schwartz:NC_gravity}, i.e.\ from what might be called a relativist's
point of view, and introduce the locally Galilei-invariant potential.
For a somewhat more `field-theoretic' perspective on the subject, we
refer to the original literature on type~\acronym{I}
\cite{Christensen.EtAl:2014a,Christensen.EtAl:2014b,
  Bergshoeff.EtAl:2015} and type~\acronym{II} \acronym{TTNC} geometry
\cite{Van_den_Bleeken:2017,Hansen.EtAl:2019,Hansen.EtAl:2020}, and in
particular the review \cite{Hartong.EtAl:2023}.

The basic geometric structure in Newton--Cartan geometry is that of a
\emph{Galilei manifold}, i.e.\ a differentiable manifold $M$ of
dimension $n+1$ ($n \ge 1$) with a nowhere-vanishing \emph{clock form}
$\tau \in \Omega^1(M)$ and a symmetric contravariant 2-tensor field
$h$, the \emph{space metric}, which is positive semidefinite of rank
$n$, such that the degenerate direction of $h$ is spanned by $\tau$.
The latter condition may be expressed as
\begin{equation}
  \tau_\mu h^{\mu\nu} = 0 \; .
\end{equation}
Vectors in the kernel of $\tau$ are called \emph{spacelike}, other
vectors are called \emph{timelike}.  The integral of $\tau$ along any
worldline (i.e.\ curve) in $M$ is interpreted as the time elapsed
along this worldline.  If $\D\tau = 0$, the time between two events
(i.e.\ points in $M$) is (locally) independent of the connecting
worldline chosen to measure it; one then speaks of a Galilei manifold
with \emph{absolute time}.  If only the weaker condition $\tau \wedge
\D\tau = 0$ holds, such that by Frobenius' theorem the distribution
$\ker\tau$ of spacelike vectors is integrable (i.e.\ we have an
absolute notion of simultaneity), then one says to be in the situation
of \emph{twistless-torsional Newton--Cartan} (\acronym{TTNC})
geometry.  In any case, $h$ induces a positive-definite bundle metric
on the spacelike vectors.  In the following, unless otherwise stated
we will always assume the `twistless torsion' condition $\tau \wedge
\D\tau = 0$.

A \emph{(local) Galilei frame} on a Galilei manifold is a local frame
$(v, \e_a)$, $a = 1, \dots, n$, of vector fields satisfying
\begin{equation}
  \tau(v) = 1, \;
  h^{\mu\nu} = \delta^{ab} \e_a^\mu \e_b^\nu \; .
\end{equation}
In particular, $v$ is a unit timelike vector field.  It follows that
$\tau(\e_a) = 0$, i.e.\ the $\e_a$ are spacelike vector fields, such
that the dual frame of one-forms takes the form $(\tau, \e^a)$.
Changes from one choice of Galilei frame $(v, \e_a)$ to another
Galilei frame $(v', \e_a')$ are called \emph{local Galilei
  transformations}; they are precisely given by basis change matrix
functions with values in the (orthochronous) homogeneous Galilei group
$\Gal = \mathsf{O}(n) \ltimes \R^n \subset \mathsf{GL}(n+1)$.
Therefore, Galilei frames may be understood as local sections of the
\emph{Galilei frame bundle} $G(M)$ of $(M,\tau,h)$, a principal
$\Gal$-bundle which is a reduction of the structure group of the
general linear frame bundle of $M$.  Local Galilei \emph{boosts}, also
called \emph{Milne boosts}, are changes of only the unit timelike
vector field $v$ of a Galilei frame $(v, \e_a)$.  These are of the
form
\begin{equation}
  \label{eq:Milne_boost}
  v \mapsto v' = v - k^a \e_a
\end{equation}
for an $\R^n$-valued boost velocity function $k$.  The dual frame
$(\tau, \e^a)$ then transforms according to
\begin{equation}
  (\tau, \e^a) \mapsto (\tau, \e^a + k^a \tau) \; .
\end{equation}

Note that we will not discuss connections compatible with the metric
structure of a Galilei manifold in this section, but only later in
\cref{sec:comp_conn}.

There is an additional ingredient that is commonly viewed as part of
the geometric structures defining a Newton--Cartan geometry,
particularly in the \acronym{TTNC} case \cite{Duval.Kuenzle:1977,
  Duval.Kuenzle:1984,Andringa.EtAl:2011,Geracie.EtAl:2015,
  Schwartz:2023}.  Locally, it is given by the specification of a
one-form $a$ for each choice of local Galilei frame in such a way that
it is invariant under spatial rotations of the frame and under local
Galilei boosts \eqref{eq:Milne_boost} transforms according to
\begin{equation}
  \label{eq:a_boost}
  a \mapsto a + \delta_{ab} k^a \e^b + \frac{1}{2} |k|^2 \tau \; .
\end{equation}
Globally, it may be understood as follows \cite{Duval.Kuenzle:1977,
  Duval.Kuenzle:1984,Schwartz:2023}.  We consider the representation
$\dot\rho \colon \Gal \to \mathsf{GL} (\R^{n+1} \oplus \ui)$ given by
\begin{equation}
  \dot\rho_{(R,k)}(y^t, y^a, \I\varphi)
  = \left(y^t, \tensor{R}{^a_b} y^b + y^t k^a,
    \I(\varphi + \tfrac{1}{2} |k|^2 y^t + k_a \tensor{R}{^a_b} y^b)
  \right) \; .
\end{equation}
On the Galilei frame bundle $G(M)$, we have a tensorial
$\R^{n+1}$-valued one-form $\btheta$ corresponding to the canonical
solder form of $G(M) \times_\Gal \R^{n+1} \cong TM$.  The global
object locally represented by the form $a$ from above is then a
one-form $\ba$ on $G(M)$ that together with this $\btheta$ yields a
$\dot\rho$-tensorial form $(\btheta,\I\ba)$.  The local form $a$ is
then the pullback of $\ba$ along the local Galilei frame $(v, \e_a)$
understood as a section of $G(M)$.  The representation $\dot\rho$
arises in the study of the Bargmann group, whose Lie algebra is (for
$n \ne 2$) the essentially unique non-trivial 1-dimensional central
extension of the inhomogeneous Galilei algebra.  For that reason, we
will call a field $\ba$ as described here a \emph{Bargmann form on
  $(M,\tau,h)$}.\footnote{\label{fn:name_a}Note that in some of the
  modern literature on Newton--Cartan geometry, the (local) form we
  denote here by $a$ is denoted by $m$ instead.  Also, at least in the
  type~\acronym{I} \acronym{TTNC} case---for more on type~\acronym{I}
  and type~\acronym{II} see below---, this field $m$ is commonly
  called the `mass gauge field/potential' or similar.  In the
  type~\acronym{II} case, however, it does not have the interpretation
  of a gauge potential related to the central `mass direction' of the
  Bargmann algebra, so using this name would be misleading.  Since the
  transformation behaviour under boosts---that is, its interpretation
  as a global object on $G(M)$---is the same in both the
  type~\acronym{I} and type~\acronym{II} cases, we wanted to use a
  common name for both cases, settling on `Bargmann form'.  In
  ref.~\cite{Schwartz:2023}, one of the present authors used the
  terminology `Bargmann structure' instead, which however (a) has been
  used with a different meaning in previous literature
  \cite{Duval.EtAl:1985} and (b) is potentially confusing because of
  the notion of $G$-structures on manifolds.}

Using a Bargmann form $\ba$ on $(M,\tau,h)$, one can construct a unit
timelike vector field $\hat v \in \Gamma(TM)$ that is \emph{locally
  Galilei-invariant}, as follows.  Let $(v, \e_a)$ be a local Galilei
frame and $a$ the pullback of $\ba$ along $(v, \e_a)$.\footnote{Since
  $a$ is invariant under spatial rotations of the local Galilei frame
  $(v, e_a)$, it really only depends on $v$.}  We then define $\hat v$
according to
\begin{equation}
  \label{eq:defn_v_hat}
  \hat v^\mu := v^\mu + h^{\mu\nu} a_\nu \; .
\end{equation}
The vector field $\hat v$ is Galilei-invariant in the following sense:
replacing $(v, \e_a)$ by any other local Galilei frame, i.e.\ applying
a local Galilei transformation to it, the vector field $\hat v$
defined by \cref{eq:defn_v_hat} will be \emph{the same}.  For Galilei
frames related by spatial frame rotations this is obvious; for Galilei
frames related by Milne boosts it follows by direct computation.  This
is what we mean by calling $\hat v$ `Galilei-invariant'.  Put
differently, this means that $\hat v$ is determined just by $\tau$,
$h$, and $\ba$.

The vector field $\hat v$ is widely considered in the literature (see,
e.g., references~\cite{Hartong.EtAl:2023,Bekaert.Morand:2016,
  Van_den_Bleeken:2017}), but rarely given a name.  In
ref.~\cite{Bekaert.Morand:2016} it is called `Coriolis-free field of
observers'.

Since $\hat v$ is itself a unit timelike vector field, we may consider
the representative of the Bargmann form $\ba$ with respect to $\hat
v$, i.e.\ the pullback of $\ba$ along any local Galilei frame whose
timelike vector field is $\hat v$.  This representative we denote by
$\hat a$.  Note that, by definition, this $\hat a$ is---just as $\hat
v$---fully determined by $\tau$, $h$, and $\ba$.

We may compute $\hat a$ as follows: for any (fixed, but abitrary) unit
timelike $v$, we may view the Galilei-invariant $\hat v$ as `just
another unit timelike vector field' that arises from $v$ by applying a
specific Milne boost, given by \eqref{eq:defn_v_hat}.  Thus, we can
compute $\hat a$ using \eqref{eq:a_boost}.  This yields the explicit
result
\begin{subequations} \label{eq:a_hat}
\begin{align}
  \hat a &= \hat\phi \tau \; \\
  \shortintertext{with}
  \label{eq:phi_hat}
  \hat\phi &= a(v) + \frac{1}{2} h(a,a) \; .
\end{align}
\end{subequations}

Since we know that $\hat a$ is determined by $\tau$, $h$, and $\ba$,
we know that the function $\hat\phi$ appearing in \eqref{eq:a_hat} is
determined by these fields as well.  Thus $\hat\phi$ is locally
Galilei-invariant in the same sense as $\hat v$: the explicit formula
\eqref{eq:phi_hat} yields the same result for any choice of local
Galilei frame $(v, \e_a)$.  (Instead of the abstract argument just
presented, this may also be checked by direct calculation.)  Note that
in the literature, no common name is used for $\hat\phi$: for example,
ref.~\cite{Bergshoeff.EtAl:2015} calls it `Newton potential'; in
ref.~\cite{Bekaert.Morand:2016} it is called `gravitational gauge
scalar'.  We call $\hat\phi$ the \emph{locally Galilei-invariant
  potential}.  (Some motivation for why we choose this name will be
given in \cref{fn:name_phi_hat} in \cref{sec:application}.)

Note that the representative $\hat a$ of $\ba$ with respect to the
locally Galilei-invariant unit timelike vector field $\hat v$ is
proportional to $\tau$.  Conversely, this may be used to \emph{define}
$\hat v$: one easily checks that, for $\tau$, $h$, $\ba$ given, $\hat
v$ is the \emph{unique} unit timelike vector field for which the local
representative of $\ba$ is proportional to $\tau$.

We wish to explicitly stress a somewhat subtle conceptual point here,
which might be prone to confusion.  When we computed $\hat a$ above,
we did so by understanding $v \mapsto \hat v$ as a Milne boost.  In
general, however, the consideration of quantities associated with
$\hat v$ does \emph{not} necessarily mean that one `fixes' the Milne
boost symmetry, i.e.\ that one fixes $\hat v$ to be that unit timelike
vector field to which local quantities depending on the choice of
local frame refer.  Instead, we still have all the freedom in the
choice of $v$.  (If this were not the case, it would clearly make no
sense at all to say that $\hat v$ be `locally Galilei-invariant'.)

We conclude this section by discussing the different kinds of `gauge
transformations' in Newton--Cartan geometry, i.e.\ transformations
relating mathematical situations which are to be considered equivalent
descriptions of `the same' geometric situation.  First, due to the
differential-geometric nature of the theory, there are
diffeomorphisms: acting on \emph{all} the involved objects with
pushforward\footnote{Given a diffeomorphism $\varphi \colon M \to N$,
  by applying its differential $\mathrm{D}\varphi$ to bases, we obtain
  a principal bundle isomorphism $\boldsymbol{\varphi} \colon G(M) \to
  G(N)$ between the Galilei frame bundles $G(M)$ of $(M,\tau,h)$ and
  $G(N)$ of $(N,\varphi_*\tau,\varphi_*h)$ in a natural way.  This
  defines what we mean by pushforward of a Bargmann form: given a
  Bargmann form $\ba$ on $(M,\tau,h)$, its pushforward
  $\boldsymbol{\varphi}_*\ba$ by this induced bundle isomorphism is
  then a Bargmann form on $(N,\varphi_*\tau,\varphi_*h)$.} by a
diffeomorphism $\varphi \colon M \to N$ leads to a different
mathematical description of the same geometric situation.  Second,
there are local Galilei transformations: these are just changes in the
(arbitrary) choice of local Galilei frame $(v,\e_a)$, giving local
descriptions of global objects living on $G(M)$; all the true global
geometric objects such as $\tau,h,\ba$ stay invariant.  When we use
the name \emph{gauge transformation} in the following, we will always
mean the third kind: certain transformations of the Bargmann form
$\ba$.  These come in two different flavours, dubbed
\emph{type~\acronym{I}} and \emph{type~\acronym{II}}, distinguishing
two different kinds of \acronym{TTNC} geometry.

In type~\acronym{I} \acronym{TTNC} geometry, the Bargmann form is
interpreted as arising from a `gauging' of the central direction of
the Bargmann algebra \cite{Andringa.EtAl:2011,
  Geracie.EtAl:2015}.\footnote{From a global point of view, this
  corresponds to the observation that a Bargmann form $\ba$ and a
  principal connection $\bomega$ on $G(M)$ together determine in a
  natural way a principal connection $\boldsymbol{\hat\omega}$ on a
  principal bundle $B(M)$ that extends $G(M)$ and whose structure
  group is the Bargmann group \cite{Duval.Kuenzle:1977,
    Duval.Kuenzle:1984,Schwartz:2023}.}  The corresponding $\Ui$ gauge
transformations locally take the form
\begin{equation}
  \label{eq:gauge_trafo_type_I}
  a \mapsto a + \D\chi
\end{equation}
for functions $\chi \in C^\infty(M)$.  Type~\acronym{I} \acronym{TTNC}
geometry naturally arises by null reduction of Lorentzian geometry
\cite{Duval.EtAl:1985,Julia.Nicolai:1995}.

Type~\acronym{II} \acronym{TTNC} geometry arises instead from a formal
expansion of Lorentzian geometry in $c^{-1}$, where $c$ is the
velocity of light \cite{Van_den_Bleeken:2017,Hansen.EtAl:2019,
  Hansen.EtAl:2020,Hartong.EtAl:2023}.\footnote{For the case of
  type~\acronym{II} \acronym{TTNC} geometry, one usually also
  considers part of the geometric structure the subleading part of the
  spacelike Lorentzian coframe fields, commonly denoted $\pi^a$, which
  however won't play a role for our considerations.}  Here, $a$ is the
next-to-leading-order part of the timelike Lorentzian coframe field.
The gauge transformations it inherits from subleading
$c^{-2}$-dependent diffeomorphisms take the form
\begin{equation}
  \label{eq:gauge_trafo_type_II}
  a \mapsto a - \mathcal L_\zeta \tau
  = a - \D\tau(\zeta, \cdot) - \D(\tau(\zeta))
\end{equation}
for vector fields $\zeta \in \Gamma(TM)$.  Note that in the case of
absolute time ($\D\tau = 0$), type~\acronym{I} and type~\acronym{II}
gauge transformations coincide.  Note also that the locally
Galilei-invariant vector field $\hat v$ \eqref{eq:defn_v_hat} is not
gauge-invariant (in both the type~\acronym{I} and type~\acronym{II}
cases).

\section{Gauge transformations of the locally Galilei-invariant
  potential}
\label{sec:gauge_phi_hat_0}

In this section, we are going to show that in both type~\acronym{I}
and type~\acronym{II} \acronym{TTNC} geometry, there is always a local
gauge transformation that transforms the locally Galilei-invariant
potential \eqref{eq:phi_hat} to zero.  Notably, our arguments will use
weaker regularity assumptions than those that are implicitly made in
typical discussions of similar gauge fixing results in the literature.

\subsection{Type~\acronym{I} \acronym{TTNC} geometry}
\label{sec:gauge_phi_hat_0_I}

Under a type~\acronym{I} \acronym{TTNC} gauge transformation
\eqref{eq:gauge_trafo_type_I} of $\ba$, the locally Galilei-invariant
potential \eqref{eq:phi_hat} transforms according to
\begin{align}
  \hat\phi
  &\mapsto (a + \D\chi)(v) + \frac{1}{2} h(a + \D\chi, a + \D\chi)
    \nonumber\\
  &= \hat\phi + \D\chi(v) + h(\D\chi, a) + \frac{1}{2} h(\D\chi,
    \D\chi) \nonumber\\
  &= \hat\phi + \D\chi(\hat v) + \frac{1}{2} h(\D\chi, \D\chi) \; .
\end{align}
Therefore, we may transform $\hat\phi$ to zero by performing a gauge
transformation with the gauge parameter $\chi$ solving the
\acronym{PDE}
\begin{equation}
  \label{eq:phi_hat_zero}
  0 = \hat\phi + \D\chi(\hat v) + \frac{1}{2} h(\D\chi, \D\chi) \; .
\end{equation}

This equation looks suspiciously like a Hamilton--Jacobi equation.
And indeed it may be brought into the form of one: due to $\tau$
satisfying the `twistless torsion' condition $\tau \wedge \D\tau = 0$,
there are local coordinates $(t,x^a)$ and a function $f$ (an
`integrating factor') such that locally $\tau = f \D t$; since $\tau$
is nowhere-vanishing, so is $f$.  Together with $\tau_\mu h^{\mu\nu} =
0$, this implies $h = h^{ab} \partial_a \otimes \partial_b$ (using the
notation $\partial_a = \frac{\partial}{\partial x^a}$), and $\tau(\hat
v) = 1$ implies $\hat v = \frac{1}{f} \partial_t + \hat v^a
\partial_a$.  Hence \eqref{eq:phi_hat_zero} takes the coordinate form
\begin{subequations}
\begin{align}
  0 &= \hat\phi + \frac{1}{f} \partial_t\chi + \hat v^a \partial_a\chi
      + \frac{1}{2} h^{ab} \partial_a\chi \partial_b\chi \; , \\
  \intertext{or equivalently}
  0 &= f \hat\phi + \partial_t\chi + f \hat v^a \partial_a\chi
      + \frac{1}{2} f h^{ab} \partial_a\chi \partial_b\chi \; .
\end{align}
\end{subequations}
This is precisely a Hamilton--Jacobi equation
\begin{subequations}
\begin{align}
  0 &= H\left(t, \vec x,
        \frac{\partial\chi(t, \vec x)}{\partial\vec x}\right)
      + \frac{\partial\chi(t, \vec x)}{\partial t} \\
  \intertext{with Hamiltonian}
  H(t, \vec x, \vec p)
    &= \frac{1}{2} f(t, \vec x) h^{ab}(t, \vec x) p_a p_b
      + f(t, \vec x) \hat v^a(t, \vec x) p_a
      + f(t, \vec x) \hat\phi(t, \vec x) \; .
\end{align}
\end{subequations}
The Hamiltonian being sufficiently regular (which is the case if
$\tau$, $h$, $\ba$, and $v$ are sufficiently regular), the
corresponding Hamiltonian equations of motion admit (local) solutions,
which is equivalent to the Hamilton--Jacobi equation admitting a local
solution. Thus, we can always perform a local gauge transformation
such that after the transformation we have $\hat\phi = 0$.

Note that demanding the fields to be `sufficiently regular' in the
above argument is a rather weak assumption: if the Hamiltonian $H$ is
$C^k$ with $k \ge 2$, locally there exist unique solutions of the
Hamiltonian equations of motion with $C^{k-1}$ dependence on the
initial conditions, with $t$ dependence also (at least) $C^{k-1}$.
This implies that locally a $C^{k-1}$ solution of the Hamilton--Jacobi
equation may be constructed (namely Hamilton's `principal function',
i.e.\ the action integral over a trajectory with fixed starting point
and variable end point).

Thus our proof is based on significantly weaker assumptions than those
typically used for similar results in the existing literature: in
ref.~\cite[prop.~3.26]{Bekaert.Morand:2016}, which treats the special
case of our result in the absolute time case $\D\tau = 0$ (see
\cref{thm:boost_NC_form_zero} below), it is explicitly assumed that
$h$, $v$, and $\hat\phi$ be analytic, such that the authors may appeal
to the Cauchy--Kovalevskaya theorem.  In literature with less explicit
emphasis on mathematical rigour, similar gauge fixing results are
typically justified by simply arguing that there are as many
differential equations as parameter functions, without discussing the
issue further; see, e.g., ref.~\cite{Van_den_Bleeken:2017} after
equation (43).  Such arguments therefore also need to appeal to some
existence result, which in the general case can only be the
Cauchy--Kovalevskaya theorem, requiring analyticity.

\subsection{Type~\acronym{II} \acronym{TTNC} geometry}
\label{sec:gauge_phi_hat_0_II}

Under a type~\acronym{II} subleading diffeomorphism
\eqref{eq:gauge_trafo_type_II} parametrised by a vector field $\zeta =
-\chi v + \lambda$ with $\lambda$ spacelike, the locally
Galilei-invariant potential $\hat\phi$ \eqref{eq:phi_hat} transforms
as
\begin{align}
  \label{eq:phi_hat_trafo_type_II}
  \hat\phi
  &\mapsto (a - \D\tau(\zeta, \cdot) + \D\chi)(v)
    + \frac{1}{2} h(a - \D\tau(\zeta, \cdot) + \D\chi,
      a - \D\tau(\zeta, \cdot) + \D\chi) \nonumber\\
  &= \hat\phi + \D\chi(\hat v) + \frac{1}{2} h(\D\chi, \D\chi)
    - \D\tau(\zeta, \hat v)
    - h(\D\tau(\zeta, \cdot), \D\chi)
    + \frac{1}{2} h(\D\tau(\zeta, \cdot), \D\tau(\zeta, \cdot)) \; .
\end{align}

Setting the spacelike part $\lambda$ to zero, this becomes
\begin{align}
  \hat\phi
  &\mapsto \hat\phi + \D\chi(\hat v) + \frac{1}{2} h(\D\chi, \D\chi)
    + \chi \D\tau(v, \hat v)
    + \chi h(\D\tau(v, \cdot), \D\chi)
    + \frac{\chi^2}{2} h(\D\tau(v, \cdot), \D\tau(v, \cdot)) \; .
\end{align}
Even when imposing the twistless torsion condition $\tau \wedge \D\tau
= 0$, the terms involving $h(\D\tau(v, \cdot), \cdot)$ will in general
not vanish.  Hence, the equation for gauging $\hat\phi$ to zero
involves terms linear and quadratic in $\chi$ (non-differentiated); in
particular, differently to the type~\acronym{I} case, it is not a
Hamilton--Jacobi equation and so we cannot easily argue for it to have
a solution.

Instead, however, we now consider the transformation
\eqref{eq:phi_hat_trafo_type_II} for $\chi = 0$, i.e.\ with $\zeta =
\lambda$ spacelike.  As in the type~\acronym{I} case we use the
twistless torsion condition $\tau \wedge \D\tau = 0$ to locally write
$\tau = f \D t$, implying $\D\tau = \partial_a f \D x^a \wedge \D t$.
We then have
\begin{equation}
  \D\tau(\zeta, \cdot)
  = \D\tau(\lambda, \cdot)
  = \lambda^a \partial_a f \D t
  = \lambda^a \frac{\partial_a f}{f} \tau \; ,
\end{equation}
such that the transformation behaviour
\eqref{eq:phi_hat_trafo_type_II} of $\hat\phi$ takes the form
\begin{equation}
  \hat \phi
  \mapsto \hat\phi - \D\tau(\lambda, \hat v)
    + \frac{1}{2} h(\D\tau(\lambda, \cdot), \D\tau(\lambda, \cdot))
  = \hat\phi - \lambda^a \frac{\partial_a f}{f} \; .
\end{equation}
Therefore, in order to transform $\hat\phi$ to zero by a subleading
spacelike diffeomorphism, the parametrising spacelike vector field
$\lambda$ has to satisfy the equation
\begin{equation}
  \hat\phi = \lambda^a \partial_a \ln f \; .
\end{equation}
If $f$ is not locally spatially constant, i.e.\ $\partial_a f \ne 0$,
this equation has a local solution: for example, we may set
\begin{equation}
  \label{eq:type_II_lambda}
  \lambda^a
  = \hat\phi \, \frac{\delta^{ab} \partial_b \ln f}
     {\sqrt{\delta^{cd} (\partial_c \ln f)(\partial_d \ln f)}} \; .
\end{equation}

If instead $f$ is locally spatially constant in some region, this
means that $\D\tau = 0$ there, such that type~\acronym{I} and type
\acronym{II} gauge transformations coincide and we can again argue as
in \cref{sec:gauge_phi_hat_0_I} for the existence of a gauge
transformation transforming $\hat\phi$ to zero.

Thus, we have shown that also in the case of type~\acronym{II}
\acronym{TTNC} geometry one can always locally find a gauge
transformation (here meaning a type~\acronym{II} subleading
diffeomorphism) transforming $\hat\phi$ to zero.

Note that $\lambda$ as given in \eqref{eq:type_II_lambda} directly
inherits the regularity of the geometric fields: if $\tau$ is $C^k$,
then $f$ may be chosen $C^k$ as well; hence, further assuming that
$h$, $\ba$, and $v$ be $C^{k-1}$, we have that $\lambda$ is $C^{k-1}$
as well.  Thus, as in the type~\acronym{I} case, our gauge fixing
works in the case of finite-degree differentiability, without the need
for assuming analyticity.

\section{Application: parametrisation of \acronym{TTNC} geometries}
\label{sec:application}

In this section, we are going to apply the special gauge
transformation derived in the previous section in order to parametrise
\acronym{TTNC} geometries by either just the space metric and a unit
timelike vector field, or a Riemannian (co-)metric and a maximum-rank
`spacelike' integrable distribution.  We will also see that our gauge
fixing is a generalisation of a classical result in Newton--Cartan
gravity.

First, let us restate the result of the previous section in the
following form:
\begin{theorem}
  \label{thm:Bargmann_gauge_v_0}
  Let $(M,\tau,h)$ be a Galilei manifold satisfying the twistless
  torsion condition $\tau \wedge \D\tau = 0$, and let
  $\boldsymbol{\tilde{a}}$ be a Bargmann form on it.  Then locally
  there exist a gauge-equivalent Bargmann form $\ba$ and a unit
  timelike vector field $v$ such that the local representative of
  $\ba$ with respect to $v$ is $a = 0$.
\end{theorem}

In the type~\acronym{I} case, this implies in particular that the
local representative $\tilde{a}$ of $\boldsymbol{\tilde{a}}$ with
respect to $v$ is closed, $\D \tilde{a} = 0$, since it differs from $a$ by a
type~\acronym{I} gauge transformation.

\begin{proof}[Proof of \cref{thm:Bargmann_gauge_v_0}]
  According to the previous section, there exists a local gauge
  transformation such that the locally Galilei-invariant potential of
  the gauge-transformed Bargmann form $\ba$ vanishes, $\hat\phi = 0$.
  Then taking $v$ to be the locally Galilei-invariant vector field
  $\hat v$ determined by $\tau, h$ and $\ba$, by construction we have
  $a = \hat a = \hat\phi \tau = 0$.  (Concretely, this means that
  choosing \emph{any} unit timelike vector vield $v'$, we define
  $v^\mu = v'^\mu + h^{\mu\nu} a'_\nu$ where $a'$ is the local
  representative of $\ba$ with respect to $v'$.)
\end{proof}

Note that the statement of \cref{thm:Bargmann_gauge_v_0} really is a
reformulation of the result of \cref{sec:gauge_phi_hat_0}: if after
the gauge transformation $\ba$ has representative $a = 0$ with respect
to $v$, this representative is proportional to $\tau$, which implies
that the proportionality factor is the locally Galilei-invariant
potential $\hat\phi$, i.e.\ in this case we have $\hat\phi = 0$.
Thus, the theorem indeed says that there is a local gauge
transformation transforming $\hat\phi$ to zero.

\subsection{Parametrisation of \acronym{TTNC} geometries up to gauge}

\Cref{thm:Bargmann_gauge_v_0} implies in particular that, fixing
$\tau$ and $h$, if we let $v$ vary over the set of all unit timelike
vector fields and consider, for each $v$, the Bargmann form $\ba$
which is defined by its representative with respect to $v$ vanishing,
we will (locally) parametrise \emph{all gauge equivalence classes} of
Bargmann forms on $(M,\tau,h)$.

We can even strip $\tau$ from the prescribed data, as follows: given a
Galilei manifold $(M,\tau,h)$ and a unit timelike vector field $v$, we
have that $\gamma := h + v \otimes v$ is non-degenerate.  Conversely,
starting with just a (positive-semidefinite) symmetric contravariant
2-tensor field $h$ of rank $n$ and a vector field $v$ such that
$\gamma = h + v \otimes v$ is non-degenerate, it is easy to see that
the conditions
\begin{equation}
  \label{eq:tau_via_h_and_v}
  \tau(v) = 1 \; , \quad \tau_\mu h^{\mu\nu} = 0
\end{equation}
then determine a unique one-form $\tau$.  We can then also express the
twistless torsion condition for the $\tau$ thus determined in terms of
$h$ and $v$: the condition $\tau \wedge \D\tau = 0$ is equivalent to
$h^{\mu\rho} h^{\sigma\nu} \partial_{[\mu} \tau_{\nu]} = 0$, which can
be rewritten as
\begin{equation}
  \label{eq:TT_via_h_and_v}
  (\gamma^{-1})_{\mu\nu} v^\nu h^{\lambda[\rho} \partial_\lambda
  h^{\sigma]\mu} = 0 \; ,
\end{equation}
where $\gamma^{-1}$ is the inverse of $\gamma = h + v \otimes v$.

Combined with \cref{thm:Bargmann_gauge_v_0}, this shows that we may
(locally) parametrise all the basic geometric objects defining a
\acronym{TTNC} geometry up to (type~\acronym{I} or type~\acronym{II})
gauge transformations by just $h$ and $v$:
\begin{theorem}
  \label{thm:param_TTNC} 
  Let $M$ be an $(n+1)$-dimensional differentiable manifold, $h$ a
  positive-semidefinite symmetric contravariant 2-tensor field of rank
  $n$ on $M$, and $v$ a vector field on $M$ such that (i)~$\gamma = h
  + v \otimes v$ is non-degenerate and (ii) \eqref{eq:TT_via_h_and_v}
  holds.  Consider the unique one-form $\tau \in \Omega^1(M)$
  satisfying \eqref{eq:tau_via_h_and_v}.  This makes $(M,\tau,h)$ into
  a Galilei manifold with $v$ a unit timelike vector field, and
  satisfies the twistless torsion condition $\tau \wedge \D\tau = 0$.
  Furthermore, we may define a Bargmann form $\ba$ on $(M,\tau,h)$ by
  demanding that its local representative with respect to $v$ vanish,
  $a = 0$.

  Locally, \emph{any} twistless-torsional Galilei manifold with
  Bargmann form is of this form, up to gauge transformations of the
  Bargmann form.  \qed
\end{theorem}

We now locally reformulate the assumptions of \cref{thm:param_TTNC}.
Let $M$ be a differentiable manifold of dimension $n + 1$ and $\Sigma
\subset TM$ an integrable distribution of rank $n$.  Furthermore, let
$\gamma$ be a positive-definite symmetric contravariant 2-tensor
field, that is, a positive-definite \emph{cometric}, on $M$.  Then
locally there exists a unique one-form $\tau$ such that
\begin{equation}
  \label{eq:tau_via_Sigma_gamma}
  \tau \wedge \D \tau = 0 \; , \quad
  \ker(\tau) = \Sigma \; , \quad \text{and} \quad
  \gamma(\tau, \tau) = 1 \; .
\end{equation}
We locally define the vector field
\begin{subequations} \label{eq:v_and_h_via_Sigma}
\begin{align}
  v &:= \gamma(\tau, \cdot) \\
  \intertext{and the contravariant 2-tensor field}
  h &:= \gamma - v \otimes v \; .
\end{align}
\end{subequations}
By construction, $h$ is symmetric and positive semidefinite of rank
$n$.  Furthermore, $\tau$ satisfies conditions
\eqref{eq:tau_via_h_and_v}, which with $\Sigma$ being integrable
implies that \eqref{eq:TT_via_h_and_v} holds.  Hence all assumptions
of \cref{thm:param_TTNC} are locally satisfied.

Thus, in terms of $\Sigma$ and $\gamma$, we may \emph{also} (locally)
parametrise all twistless-torsional Galilei manifolds with Bargmann
form up to gauge:
\begin{theorem}
  \label{thm:param_TTNC_Sigma}
  Let $M$ be an $(n+1)$-dimensional differentiable manifold, $\Sigma
  \subset TM$ an integrable distribution of rank $n$, and $\gamma$ a
  positive-definite cometric on $M$.  Consider the unique local
  one-form $\tau$ on $M$ satisfying \eqref{eq:tau_via_Sigma_gamma},
  and define $v$ and $h$ by \eqref{eq:v_and_h_via_Sigma}.  This
  locally makes $(M,\tau,h)$ into a Galilei manifold with $v$ a unit
  timelike vector field, and satisfies the twistless torsion condition
  $\tau \wedge \D\tau = 0$.  Furthermore, we may define a Bargmann
  form $\ba$ on $(M,\tau,h)$ by demanding that its local
  representative with respect to $v$ vanish, $a = 0$.

  Locally, \emph{any} twistless-torsional Galilei manifold with
  Bargmann form is of this form, up to gauge transformations of the
  Bargmann form.  \qed
\end{theorem}

\subsection{Interlude: compatible connections}
\label{sec:comp_conn}

Now we are going to discuss, in addition to the basic metric structure
of a Galilei manifold (and possibly a Bargmann form on it), compatible
connections.  For details on this material we refer to
references~\cite{Geracie.EtAl:2015,Schwartz:2023,Schwartz:2025}; a
more general discussion may be found in
ref.~\cite{Bergshoeff.EtAl:2023}.

A connection compatible with the structure of a Galilei manifold
$(M,\tau,h)$ is called a \emph{Galilei connection}.  Phrased in terms
of a covariant derivative operator, this amounts to a covariant
derivative $\nabla$ on the tangent bundle $TM$ compatible with $\tau$
and $h$, i.e.\ satisfying
\begin{equation}
  \nabla\tau = 0 \; , \quad \nabla h = 0 \; .
\end{equation}
Compatibility with $\tau$ implies that the torsion $T$ of any Galilei
connection satisfies $\tau_\rho \tensor{T}{^\rho_{\mu\nu}} =
(\D\tau)_{\mu\nu}$.  A Galilei connection on $(M,\tau,h)$ may also be
understood as a principal connection $\bomega$ on the Galilei frame
bundle $G(M)$ of $(M,\tau,h)$.  Its local connection form with respect
to a Galilei frame $(v,\e_a)$ is then a locally defined one-form
$(\tensor{\omega}{^a_b}, \varpi^a)$ taking values in the Galilei
algebra $\mathfrak{gal} = \mathfrak{so}(n) \oright \R^n$, with
rotational part $\tensor{\omega}{^a_b}$ and boost part $\varpi^a$.

Due to the degeneracy of the metric structure, differently to the
pseudo-Riemannian case Galilei connections are not uniquely determined
by their torsion.  Instead, with respect to a choice of unit timelike
vector field $v$, a Galilei connection $\nabla$ is uniquely determined
by its torsion $T$ and its \emph{Newton--Coriolis form} $\Omega$ with
respect to $v$.  The latter is a two-form that may be written as
\begin{equation}
  \Omega_{\mu\nu} = 2(\nabla_{[\mu} v^\rho) h_{\nu]\rho} \; ,
\end{equation}
where $h_{\mu\nu}$ are the components of the covariant space metric
with respect to $v$, which is defined by $h_{\mu\nu} = h_{\nu\mu}$,
$h_{\mu\nu} v^\nu = 0$, $h_{\mu\nu} h^{\nu\rho} = \delta_\mu^\rho -
v^\rho \tau_\mu$.  Alternatively, extending $v$ to a local Galilei
frame $(v,\e_a)$, $\Omega$ can be written in terms of the boost part
of the local connection form and the dual frame as
\begin{equation}
  \Omega = \delta_{ab} \, \varpi^a \wedge \e^b \; .
\end{equation}
Conversely, choosing an arbitrary tensor field $T$ subject to the
constraints $\tensor{T}{^\rho_{\mu\nu}} = -\tensor{T}{^\rho_{\nu\mu}}$
and $\tau_\rho \tensor{T}{^\rho_{\mu\nu}} = (\D\tau)_{\mu\nu}$, a unit
timelike vector field $v$, and an arbitrary two-form $\Omega$, there
is a (unique) Galilei connection whose torsion is $T$ and whose
Newton--Coriolis form with respect to $v$ is $\Omega$.

By definition, together with the tensorial $\R^{n+1}$-valued one-form
$\btheta$ on $G(M)$ which corresponds to the canonical solder form of
$G(M) \times_\Gal \R^{n+1} \cong TM$, any Bargmann form $\ba$ combines
to a tensorial $(\R^{n+1} \oplus \ui)$-valued form $(\btheta,\I\ba)$
on $G(M)$.  The exterior covariant derivative
$\D^\bomega(\btheta,\I\ba)$ of this tensorial form with respect to a
Galilei connection $\bomega$ is the so-called \emph{extended torsion
  of $\bomega$ with respect to $\ba$}.  The local representative of
the extended torsion with respect to a local Galilei frame $\sigma =
(v,\e_a)$ may be seen to be given by
\begin{equation}
  \sigma^*\D^\bomega(\btheta,\I\ba) = (T^A, \I(\D a + \Omega)) \; \text{:}
\end{equation}
it consists of the frame components $T^A$ of the torsion and the
so-called \emph{mass torsion} $\D a + \Omega$.  In particular, we see
that knowing a Galilei connection's extended torsion with respect to
some Bargmann form $\ba$ amounts to knowing its torsion and
Newton--Coriolis form.  Hence, fixing a Bargmann form, \emph{Galilei
  connections are uniquely determined by their extended torsion}.

In standard Newton--Cartan gravity, Newtonian gravity is encoded by
so-called \emph{Newtonian} connections.  These are torsion-free
Galilei connections (on a Galilei manifold with absolute time) whose
curvature tensor satisfies the symmetry condition
${\tensor{R}{^\mu_\rho^\nu_\sigma}=\tensor{R}{^\nu_\sigma^\mu_\rho}}$
(where the third index was raised with $h$).  This is equivalent to
the Newton--Coriolis form (with respect to any $v$) being closed,
$\D\Omega = 0$ (this equivalence is not obvious, but requires a
somewhat lengthy calculation).\footnote{\label{fn:name_phi_hat}We may
  now explain the inspiration for the name `locally Galilei-invariant
  potential' for $\hat\phi$.  In standard Newton--Cartan gravity (with
  absolute time, $\D\tau = 0$), the Newton--Coriolis form of a
  (torsion-free) Galilei connection $\nabla$ with respect to any unit
  timelike vector field $v$ is given by $\Omega = \tau \wedge \alpha +
  2\omega$, where $\alpha$ and $\omega$ are the acceleration and twist
  of $v$ with respect to $\nabla$, respectively.  For a vector field
  $v$ with vanishing $\omega$, the connection being Newtonian, i.e.\
  $\D\Omega = 0$, is then equivalent to $0 = \tau \wedge \D\alpha$,
  which by the Poincaré lemma is locally equivalent to $\alpha =
  \D\phi - v(\phi) \tau$ for some function $\phi$.  Additionally
  assuming spatial flatness and rigidness of $v$ (i.e.\ $\mathcal L_vh
  = 0$)---which together with vanishing twist implies that $\nabla$
  satisfies Trautman's \emph{absolute rotation} condition
  $\tensor{R}{^{\mu\nu}_{\rho\sigma}} = 0$---we recover the standard
  formulation of Newtonian gravity in adapted coordinates with respect
  to $v$, with this function $\phi$ playing the role of the Newtonian
  potential.

  Due to $\nabla$ being Newtonian, we may (at least locally) demand it
  to be the unique extended-torsion-free Galilei connection with
  respect to $\ba$, which as an equation takes the form $\Omega = -\D
  a$ and therefore determines $\ba$ up to (type~\acronym{I}) gauge
  transformations.  In the case discussed above, we have $\Omega =
  \tau \wedge \D\phi = -\D(\phi \tau)$, so we may take $a = \phi
  \tau$.  So in this special situation and for this choice of $a$, the
  proportionality factor between $a$ and $\tau$ plays the role of the
  Newtonian potential, which is why we call the locally
  Galilei-invariant quantity $\hat\phi$ from \eqref{eq:phi_hat} the
  locally Galilei-invariant \emph{potential}.  Note however that after
  performing a gauge transformation, the new $\hat\phi$ will no longer
  be equal to the Newtonian potential $\phi$ with respect to $v$.}

By the Poincaré lemma, a torsion-free Galilei connection being
Newtonian is locally equivalent to the Newton--Coriolis form being
exact, $\Omega = -\D a$ for some one-form $a$.  This means that
locally, Newtonian connections are precisely those Galilei connections
whose extended torsion with respect to some Bargmann form vanishes.

\subsection{Parametrisation of standard Newton--Cartan geometry}

We can now see how our results imply the following classical result
from standard Newton--Cartan geometry \cite[thm.~3.6]
{Dombrowski.Horneffer:1964}, \cite[prop.~4.3.7]{Malament:2012},
\cite[prop.~3.26]{Bekaert.Morand:2016}:
\begin{corollary}
  \label{thm:boost_NC_form_zero}
  Let $(M,\tau,h)$ be a Galilei manifold with absolute time, $\D\tau =
  0$, and let $\bomega$ be a Newtonian connection on it.  Then locally
  there exists a unit timelike vector field $v$ such that the
  Newton--Coriolis form of $\bomega$ with respect to $v$ vanishes.

  \begin{proof}
    Locally there exists a Bargmann form $\boldsymbol{\tilde{a}}$ such
    that $\bomega$ is the (unique) Galilei connection whose extended
    torsion with respect to $\boldsymbol{\tilde{a}}$ vanishes.  By
    \cref{thm:Bargmann_gauge_v_0}, locally there exists a unit
    timelike vector field $v$ such that the local representative
    $\tilde{a}$ of $\boldsymbol{\tilde{a}}$ with respect to $v$ is
    closed.  Hence, the extended torsion vanishing implies $0 =
    \D\tilde{a} + \Omega = \Omega$.\looseness-1
  \end{proof}
\end{corollary}

We see that \cref{thm:Bargmann_gauge_v_0} is the natural
generalisation of this result to the \acronym{TTNC} case.

Note that while Malament's proof of \cref{thm:boost_NC_form_zero}
\cite[prop.~4.3.7]{Malament:2012} is very different in spirit from our
approach, the original proof by Dombrowski and Horneffer
\cite[thm.~3.6]{Dombrowski.Horneffer:1964} is very close (at least to
this special case of our gauge fixing): it also reduces the statement
(which is given in an appreciably different formulation) to solving a
Hamilton--Jacobi equation.  Bekaert and Morand prove the result (again
in a somewhat different formulation) in essentially the same way
\cite[prop.~3.26]{Bekaert.Morand:2016}, however without realising that
the \acronym{PDE} involved is a Hamilton--Jacobi equation, and
therefore requiring analyticity.\footnote{Geracie et
  al.~\cite{Geracie.EtAl:2015} claim, without proof, that the
  statement of \cref{thm:boost_NC_form_zero} follows from the
  transformation behaviour of $\Omega$ under local Galilei boosts.
  However, fleshing out this argument, one realises that it also
  amounts to solving a Hamilton--Jacobi equation.}

Using \cref{thm:boost_NC_form_zero}, we may (locally) parametrise,
similar to \cref{thm:param_TTNC}, absolute-time Galilei manifolds with
Newtonian connections in terms of just $h$ and $v$.  Similar to
\eqref{eq:TT_via_h_and_v}, the absolute time condition $\D\tau = 0$ is
equivalent to
\begin{equation}
  \label{eq:AT_via_h_and_v}
  \partial_{[\mu} ((\gamma^{-1})_{\nu]\rho} v^\rho) = 0 \; .
\end{equation}
\begin{theorem}
  \label{thm:param_ATNC}
  Let $M$ be an $(n+1)$-dimensional differentiable manifold, $h$ a
  positive-semidefinite symmetric contravariant 2-tensor field of rank
  $n$ on $M$, and $v$ a vector field on $M$ such that (i)~$\gamma = h
  + v \otimes v$ is non-degenerate and (ii)~\eqref{eq:AT_via_h_and_v}
  holds.  Consider the unique one-form $\tau \in \Omega^1(M)$
  satisfying \eqref{eq:tau_via_h_and_v}.  This makes $(M,\tau,h)$ into
  a Galilei manifold with absolute time, and $v$ a unit timelike
  vector field.  Furthermore, we may define a Newtonian connection on
  $(M,\tau,h)$ by demanding that its torsion and Newton--Coriolis form
  with respect to $v$ vanish, $\Omega = 0$.

  Locally, \emph{any} Galilei manifold with absolute time and
  Newtonian connection is of this form.  \qed
\end{theorem}

As for \cref{thm:param_TTNC}, we can locally reformulate the
assumptions for \cref{thm:param_ATNC}, yielding a reformulation
similar to \cref{thm:param_TTNC_Sigma}:
\begin{theorem}
  Let $M$ be an $(n+1)$-dimensional differentiable manifold, $\Sigma
  \subset TM$ an integrable distribution of rank $n$, and $\gamma$ a
  positive-definite cometric on $M$, such that the local one-form
  $\tau$ defined by \eqref{eq:tau_via_Sigma_gamma} is closed.
  Consider $v$ and $h$ defined by \eqref{eq:v_and_h_via_Sigma}.  This
  locally makes $(M,\tau,h)$ into a Galilei manifold with absolute
  time, and $v$ a unit timelike vector field.  Furthermore, we may
  define a Newtonian connection on $(M,\tau,h)$ by demanding that its
  torsion and Newton--Coriolis form with respect to $v$ vanish,
  $\Omega = 0$.

  Locally, \emph{any} Galilei manifold with absolute time and
  Newtonian connection is of this form.  \qed
\end{theorem}

\section{Conclusion}
\label{sec:conclusion}

In this paper, we have shown that in twistless-torsional
Newton--Cartan geometry (both its type~\acronym{I} and
type~\acronym{II} incarnations), there always exists a local gauge
transformation transforming the locally Galilei-invariant potential to
zero.  Equivalently, there exist (locally) a gauge transformation and
a unit timelike vector field such that the representative of the
gauge-transformed Bargmann form with respect to this vector field
vanishes.  We have presented the derivation of this result in full
rigour, emphasising that the assumption of a finite degree of
differentiability of the geometric fields is sufficient.

Our result generalises the classical result in standard Newton--Cartan
geometry that for any Newtonian connection, locally there exists a
unit timelike vector field that is geodesic and twist-free with
respect to the connection, i.e.\ such that the Newton--Coriolis form
of the connection with respect to this vector field vanishes.

The above result also allowed us to argue that all of the geometric
data determining a \acronym{TTNC} geometry up to gauge may be
(locally) parametrised by either just the space metric and a unit
timelike vector field, or the spacelike distribution and a
positive-definite cometric.

This possibility of `re-packaging' of all the geometric data in the
form of a distribution and a positive-definite cometric enables the
following interesting application.  The field equations of standard
Newton--Cartan gravity, as well as those of \acronym{TTNC}
(type~\acronym{II}) gravity that arise from expanding the Einstein
equations in $c^{-1}$ \cite{Van_den_Bleeken:2017}, are symmetric
covariant 2-tensor equations.  This means that the cometric is
non-degenerate and (tensorially) dual to the type of the field
equations.  Furthermore, with fixed distribution of spacelike vectors,
the cometric fully encodes the possible configurations of the system
under consideration (\acronym{TTNC} geometry).  This allows to study
application of the \emph{canonical variational completion} formalism
\cite{Voicu.Krupka:2015,Hohmann.EtAl:2021} to \acronym{TTNC} gravity.
This offers a new route towards action formulations of \acronym{TTNC}
gravity, starting from the theory (defined by its geometry and field
equations) alone, thus complementing existing approaches based on
expansions of \acronym{GR} or on symmetry considerations
\cite{Hansen.EtAl:2019, Hansen.EtAl:2020}.  We will publish these
results in upcoming work.

\section*{Acknowledgements}

We wish to thank Domenico Giulini for helpful discussions.

\printbibliography

\end{document}
